\documentclass{article}

\usepackage{amssymb,amsfonts,amsmath,stmaryrd}
\usepackage{cite,enumerate,float,indentfirst}
\usepackage{color}
\usepackage{tikz}
\usetikzlibrary{arrows,snakes,backgrounds}
\usepackage{ytableau}
\usepackage[vcentermath]{youngtab}

\def\be{\begin{eqnarray}}
\def\ee{\end{eqnarray}}
\def\nn{\nonumber}

\def\p{\partial}
\def\tr{{\rm tr}\,}
\def\Tr{{\rm Tr}\,}

\def\s{{\rm Schur}}
\def\M{\hbox{Mom}}

\definecolor{red}{rgb}{1,0,0}
\definecolor{orange}{rgb}{1,0.5,0}
\definecolor{violet}{rgb}{0.7,0,1}

\hoffset= - 1.0in         

\begin{document}

\begin{center}
\begin{small}
\hfill FIAN/TD-07/21\\
\hfill IITP/TH-10/21\\
\hfill ITEP/TH-13/21\\
\hfill MIPT/TH-09/21\\
\end{small}
\end{center}

\vspace{.5cm}

\begin{center}
\begin{Large}\fontfamily{cmss}
\fontsize{17pt}{27pt}
\selectfont
	\textbf{Matrix model partition function by a single constraint}
	\end{Large}
	
\bigskip \bigskip

\begin{large}A. Mironov$^{a,b,c,}$\footnote{mironov@lpi.ru; mironov@itep.ru},
V. Mishnyakov$^{d,a,b,}$\footnote{mishnyakovvv@gmial.com},
A. Morozov$^{d,b,c,}$\footnote{morozov@itep.ru},
R. Rashkov$^{e,f,}$\footnote{rash@phys.uni-sofia.bg; rash@hep.itp.tuwien.ac.at}
 \end{large}
\\
\bigskip

\begin{small}
$^a$ {\it Lebedev Physics Institute, Moscow 119991, Russia}\\
$^b$ {\it ITEP, Moscow 117218, Russia}\\
$^c$ {\it Institute for Information Transmission Problems, Moscow 127994, Russia}\\
$^d$ {\it MIPT, Dolgoprudny, 141701, Russia}\\
$^e$ {\it Department of Physics, Sofia University,
	5 J. Bourchier Blvd., 1164 Sofia, Bulgaria} \\
$^f$ {\it ITP, Vienna University of Technology,
	Wiedner Hauptstr. 8–10, 1040 Vienna, Austria   }
\end{small}
 \end{center}

\bigskip

\begin{abstract}
In the recent study of Virasoro action on characters,
we discovered that it gets especially simple for peculiar
linear combinations of the Virasoro operators:
particular harmonics of $\hat w$-operators.
In this letter, we demonstrate that even more is true:
a {\it single} $w$-constraint is sufficient to uniquely
specify the partition functions provided one assumes that
it is a power series in time-variables.
This substitutes the previous specifications in terms
of {\it two} requirements:
either a string equation imposed on the KP/Toda $\tau$-function
or a pair of Virasoro generators.
This mysterious {\it single}-entry definition holds
for a variety of theories, including Hermitian and complex matrix models,
and also matrix models with external matrix: the unitary and
cubic Kontsevich  models.
In these cases, it is equivalent to W-representation
and is closely related to  {\it super}\,integrability.
However, a similar single equation that completely
determines the partition function exists also in the case of the
generalized Kontsevich model (GKM) with potential of higher degree,
when the constraint algebra is a larger $W$-algebra,
and neither W-representation, nor superintegrability are
understood well enough.
\end{abstract}

\bigskip


\section{Introduction}

Matrix models \cite{Wigner,Dyson,Mehta}
is an old and respectable subject in theoretical physics,
primarily because they provide simplest solvable examples of
quantum field theory, i.e. play the role which harmonic oscillator
plays for ordinary quantum mechanics.
Though slowly, ideas from this field are also penetrating into the
field of statistics and many sciences which use ideas
of statistical distributions.
Specifics of matrix model theory is that it studies distributions
of {\it eigenvalues} and appears closely related to group theory and integrable systems.
Still, as quantum theories defined by a finite-dimensional (and well-defined)
counterpart of a path integral, matrix models possess a description
in terms of Ward identities (often referred to as
{\it Virasoro constraints} \cite{Vircon}),
which are often rich enough to define them unambiguously.
We refer to \cite{UFN3} for a comprehensive review of these
aspects of theory of matrix models, and
to \cite{superint,QKon} for an even more advanced description in terms
of {\it superintegrability}.

As old as the formal theory of matrix models is the puzzle
of interplay between (super)integrability and Ward identities.
The standard statement is that exact (non-perturbative) partition functions are
KP/Toda $\tau$-functions, but not of a generic type:
they are restricted by a single {\it string equation},
which is the lowest and the simplest of all Virasoro constraints \cite{FKN,GMMMO}.
Recently, in study \cite{MMMR1} of relation to {\it super}integrability,
we revisited action of the Virasoro operators on {\it characters}
(Schur functions), and found that especially simple is not only
action of the string equation, but also that of a $w$-constraint,
which is a peculiar linear combination of infinitely many Virasoro constraints.
In the present paper, we report a far more powerful statement:
in fact, for a matrix model partition function that satisfies $W$-constraints,
one consider instead a system of $w$-constraints that are linear combinations of all these $W$-constraints
multiplied by time variables, and
{\bf the lowest of these $w$-constraints is enough to define the partition function uniquely}, hence
neither string equation, nor integrability is needed in this approach. To put differently,
a single equation is enough to obtain the partition function,
which so far looked to be in a seeming contradiction with other standard
beliefs that the partition function is uniquely fixed either by the set of Ward identities, or by the string equation
combined with integrable equations.
In \cite{MMMR1}, this fact was overlooked, because in the Hermitian matrix model
analyzed in that reference the needed operator is actually $\hat w_{-2}$ rather
than $\hat w_{-1}$, which we considered there.
In this paper, we improve the statement and extend it to a variety of other theories,
which makes it a kind of a new basic property of matrix models.
It is intimately related to another puzzling issue, the $W$-representations
for matrix model partition functions \cite{wrep,wrep2,MSh,Alex1,Alex2}.

Given the importance of this result, we devote this whole letter to
explaining how and when it works,
leaving a less clear discussion and relation to older approaches
to another paper. The letter is organized as follows.
In the next section, we briefly remind a traditional way of
dealing with the Virasoro constraints in matrix models. In section 3,
we explain our main point in the simplest example of rectangular complex matrix model,
and, in section 4, we consider a series of examples, which includes Hermitian matrix model (where we reproduce the recent result of \cite{Max2}), and matrix models depending on external matrix: Kontsevich model, Br\'ezin-Gross-Witten model,
and, finally, the generalized Kontsevich model (GKM), where the constraint algebra is extended to
the $W$-algebra. Despite this extension, even the GKM partition function is unambiguously
encoded by a single equation.

\section{Old approach: Hermitian matrix model}

\subsection{Virasoro constraints}

The partition function of the Gaussian Hermitian matrix model is given by the integral over $N\times N$
Hermitian matrices $X$
\be\label{Herm}
Z_N\{p\} := \int \exp\left(\sum_{k=1}^\infty \frac{p_k\,\tr X^k}{k}\right)\cdot e^{-\frac{\mu}{2}\,\tr X^2} dX
\ee
understood as a power series in time-variables\footnote{These variables play the role of times in integrable KP/Toda hierarchies within the integrable framework, hence the name.} $p_k$ with the Haar measure $dX$ normalized so that $Z_N\{0\}=1$.
This partition function satisfies constraints \cite{Vircon}
\be
\hat{ L}_n^H  Z_N\{p\} = 0, \ \ \ \ n\geq -1
\label{vircon}
\ee
\be
\hat { L}_n^H := \sum_{k} (k+n)p_k\frac{\p}{\p p_{k+n}} + \sum_{a=1}^{n-1} a(n-a)\frac{\p^2}{\p p_a\p p_{n-a}}
+ 2Nn\frac{\p}{\p p_n}    + N^2\delta_{n,0} + Np_1 \delta_{n+1,0}
- \mu (n+2)\frac{\p}{\p p_{n+2}}
\label{Vir}
\ee
These operators $\hat { L}_n^H$ form a Borel subalgebra of the Virasoro algebra.

The set of constraints (\ref{vircon}) (=Ward identities, =loop equations) uniquely defines the partition function $Z_N\{p\}$ \cite{AMM,Max1}\footnote{The space of solutions to the Virasoro constraints in the general case is discussed also in \cite{AMM,AMMmore,Max2}.}. Hence, one can just solve them iteratively (see sec.\ref{solving} below), or to use the fact that the whole Borel subalgebra of the Virasoro algebra is generated just by the two constraints $L_{-1}^H$ and $L_2^H$. Another possibility is to distinguish a special role \cite{FKN} of the lowest constraint $L_{-1}^H$ called {\bf string equation} \cite{streq}: this constraint fixes a unique solution to the integrable hierarchy of the Toda chain \cite{GMMMO}. We explain this approach in the next subsection.

\subsection{String equation and integrability}

As soon as the partition function (\ref{Herm}) is a power series in $p_k$, one can present it in terms of expansion in the Schur functions $\s_R\{p_k\}$,
\be\label{chaexpH}
Z_N\{p\} = \sum_R \mu^{-|R|/2}c_R\cdot \s_R\{p\}
\ee
and, for the sake of simplicity, we put $\mu=1$ below, since it is easily restored by the grading.

Now note that there is a determinant formula for the partition function (\ref{ZH}) with an arbitrary (not obligatory Gaussian) measure $\rho(x)$ \cite{UFN3}, which provides a generic forced Toda chain solution \cite{KMMOZ,versus}:
\be
Z^{(N)}_\rho\{p\} := \int \exp\left(\sum_{k=1}^\infty \frac{p_k\,\tr X^k}{k}\right)\cdot \rho(X)dX=\det_{1\le i,j\le N}\M_{i+j-2}\{p_k\}=(-1)^{N(N-1)\over 2}\det_{1\le i,j\le N}\M_{N-i+j-1}\{p_k\}
\label{ZH}
\ee
where $\M_n\{p_k\}$ is the moment matrix
\be
\M_n\{p_k\}:=\int \exp\left(\sum_{k=1}^\infty \frac{p_k\,x^k}{k}\right)\cdot x^n\rho(x)dx
\ee
This solution is parameterized by an arbitrary function $\rho(x)$, which may be fixed by the string equation.

Now expanding the exponential in this formula, one obtains
\be
Z^{(N)}_\rho\{p\}=(-1)^{N(N-1)\over 2}\det_{1\le i,j\le N}\sum_k h_k\{p_k\}\M_{N-i+j+k-1}\{0\}=\nn\\
=(-1)^{N(N-1)\over 2}\det_{1\le i,j\le N}\sum_k h_{k+i}\{p_k\}\M_{N+j+k-1}\{0\}\stackrel{CB}{=}\nn\\
\stackrel{CB}{=}(-1)^{N(N-1)\over 2}\sum_{k_1>k_2>\ldots>k_N}
\det_{1\le i,j\le N}h_{k_i+j}\{p_k\}\cdot\det_{1\le i,j\le N}\M_{N+j+k_i-1}\{0\}=\nn\\
=(-1)^{N(N-1)\over 2}\sum_{R:\ l_R\le N}\det_{1\le i,j\le N}\M_{N-i+j+R_i-1}\{0\}\cdot \s_R\{p_k\}
\ee
where  $h_k\{p_k\}$'s are the complete homogeneous symmetric polynomials\footnote{These polynomials are defined as
\be\label{h}
\exp\Big({p_kz^k\over k}\Big)=\sum_m{h_m\{p_k\}z^m}
\ee} defined to vanish at negative values of $k$, and we used the Cauchy-Binet formula (CB), and, at the last transition, changed the variables $k_i:=R_i-i$, which allows us to have a sum over partitions $R$. We also used the first Jacobi-Trudi identity,
\be\label{JT}
\s_R\{p_k\}=\det_{i,j}h_{R_i-i+j}\{p_k\}
\ee
The normalized correlators are equal to
\be\label{cRi}
\boxed{
c_R={\det_{1\le i,j\le N} \M_{N-i+j+R_i-1}\{0\}\over \det_{1\le i,j\le N} \M_{N-i+j-1}\{0\}}
}
\ee
Now one can find it using the string equation, i.e. the Virasoro $L_{-1}^H$-constraint. In terms of the expansion coefficients $c_R$, the string equation look like \cite{MMMR1}
\be\label{s1}
\sum_\Box c_{R+\Box}=\sum_{\Box}(N-i_\Box + j_\Box)c_{R-\Box}
\ee
Note that this string equation at $N=1$ gives
\be
\M_{r+1}=r\M_{r-1}
\ee
and $\M_1=0$. This immediately gives
\be
\M_{2r}=(2r-1)!!\cdot\M_0\ \ \ \ \ \ \ \ \ \ \M_{2r-1}=0
\ee
which coincides with the moments of the Gaussian measure and demonstrates that the string equation, indeed, fixes the solution to the integrable hierarchy uniquely.

Moreover, one can calculate $c_R$ now basing on the identity for determinants:
\be\label{detF}
\det_{1\le i,j\le N}{\delta^{(2)}_{N-1+R_i-i+j}\cdot(N-2+R_i-i+j)!!}=(-1)^{N(N-1)\over 2}
\prod_{i=1}^N{(N+R_i-i)!(i-1)!\over (N-i)!}\cdot\det_{1\le i,j\le N}{\delta^{(2)}_{R_i-i+j}\over (R_i-i+j)!!}
\ee
where $\delta^{(2)}_k:=(1+(-1)^k)/2$. Now, using that\footnote{In order to derive (\ref{det}), we use (\ref{h}),
$$
\exp\Big({z^2\over 2}\Big)=\sum_k{z^{2k}\over (2k)!!}
$$
and the first Jacobi-Trudi identity (\ref{JT}).}
\be
{\s_R\{N\}\over \s_R\{\delta_{k,1}\}}=\prod_{i,j\in R}(N-i+j)=\prod_{i=1}^{N}{(N+R_i-i)!\over (N-i)!}\label{rat}\\
\s_R\{\delta_{k,2}\}=\det_{1\le i,j\le N}{\delta^{(2)}_{R_i-i+j}\over (R_i-i+j)!!}\label{det}\\
\det_{1\le i,j\le N}{\delta^{(2)}_{N-1-i+j}\cdot (N-2-i+j)!!}=(-1)^{N(N-1)\over 2}\prod_{i=1}^{N-1}i!
\ee
where we again used the first Jacobi-Trudi identity (\ref{JT}),
we finally obtain from (\ref{cRi}) formula \cite{MM}
\be
c_R = \frac{\s_R\{N\}\cdot\s_R\{\delta_{k,2}\}}{  \s_R\{\delta_{k,1}\}}
\label{cR}
\ee

\subsection{Solving Virasoro constraints iteratively\label{solving}}

Instead of using the combination of integrability and the string equation, one can solve the Virasoro constraints (\ref{vircon}) iteratively.
To this end, let us restrict the partition function to a polynomial of grading $2P$,
\be\label{Zcut}
Z_N\{p_k\}=1+\sum_{n=1}^{2P}\sum_{R\vdash n}c_R\cdot {\rm Schur}_R
\ee
i.e. cut at an arbitrary level $P$.
Then, the following claim is correct: the set of equations
\be
\left(L_n^HZ_N\{p_k\}\right)_i=0,\ \ \ \ \ i=0\ldots 2P-2-n,\ \ \ \ \ \ n=-1\ldots P-1
\ee
where $\left(L_n^HZ_N\{p_k\}\right)_i$ denotes the $i$-th graded terms in the action of the Virasoro algebra,
determines all $c_R$ but just one at the top level $P$.
The exception is the case of $P=1$, when these equations are enough to fix $c_{[2]}$ and $c_{[1,1]}$ unambiguously.

This approach requires solving a system of linear equations. In the remaining part of the paper, we demonstrate that one can instead solve a single equation, which, in the Hermitian model case, has the form
\be\label{seqH}
\boxed{\sum_{k\ge 1}p_kL^H_{k-2}Z_N\{p_k\}=0}
\ee
moreover, that this equation uniquely fixes all $c_R$ with $|R|\le 2P$ in (\ref{Zcut}) upon restricting the l.h.s. of (\ref{seqH}) to the gradings up to $2P$. This is due to emergency of recurrent relations determining a single solution.

\section{New approach: rectangular complex model}

We now start with a simpler example of the rectangular complex matrix model, and consider more examples in the next section. Our main message is that, for each model, {\bf there is a single equation that encodes the whole set of Virasoro constraints}.

\subsection{Single equation for complex model}

Partition function of the rectangular complex matrix model is given by the integral over $N_1\times N_2$ rectangular matrix $X$, \cite{comp,AMMUC}
\be\label{Ccor}
Z_{N_1,N_2}\{p_k\}:=\exp\left(\sum_{k=1}^\infty \frac{p_k\,\tr (X\bar X)^k}{k}\right)\cdot
\exp\left(-\Tr X\bar X\right) d^2X
\ee
This partition function satisfies the set of Virasoro constraints
\be\label{VirConC}
\hat L_n^C\cdot Z_{N_1,N_2}\{p_k\}=0,\ \ \ \ \ \ \ n\ge 0
\ee
\be\label{VirC}
\hat L_n^C:=\sum_k (k+n)p_k{\partial\over\partial p_{k+n}}+\sum_{a=1}^{n-1}a(n-a){\partial^2 \over \partial p_a\partial p_{n-a}}+
n(N_1+N_2){\partial\over\partial p_n}+N_1N_2\delta_{n,0}-\underline{(n+1){\partial\over\partial p_{n+1}}}
\ee
This infinite set of constraints has a unique solution, and is equivalent to a single equation
\be\label{seqC}
\boxed{
\sum_{k\ge 1} p_k \hat{L}_{k-1}^CZ_{N_1,N_2}\{p_k\}=0
}
\ee
Indeed, (\ref{seqC}) is an evident consequence of (\ref{VirConC}), and, as we demonstrate further, (\ref{seqC}) has a unique solution, similarly to (\ref{VirConC}). Hence, they are equivalent.

In order to demonstrate the uniqueness of solution to (\ref{seqC}), we rewrite it in the form
\begin{equation}\label{wn}
    \left(\hat w_{-1} + (N_1+N_2)\hat l_{-1} +N_1 N_2 p_1 - \underline{\hat l_0 }\right)Z_{N_1,N_2}\{p_k\} =0
\end{equation}
where $\hat l_n$ denote a modified Virasoro algebra:
\begin{equation}
\hat l_n:=\sum (k+n)p_{k} \frac{\partial}{\partial p_{k+n}}+\sum_{a=1}^{n-1}a(n-a) \frac{\partial^{2}}{\partial p_{a} \partial p_{n-a}}
\end{equation}
and $\hat w_m$, a part of the $W$-algebra
\begin{equation}
\hat w_{m}=\sum_{n=1} p_n \hat l_{n+m}
\end{equation}
In fact, one can consider other $\hat w$-constraints that follows from (\ref{VirConC})-(\ref{VirC}),
\be
\Big[\hat w_m+(N_1+N_2)\sum_n (m+n)p_n{\partial\over\partial p_{n+m}}+N_1N_2p_1\delta_{m,-1}
-\sum_n(n+m+1)p_n{\partial\over\partial p_{n+m+1}}\Big]Z_{N_1,N_2}\{p_k\}=0,\nn\\ m\ge -1
\ee
however, any single of these constraints but that at $m=-1$ is not enough to unambiguously fix the solution.

Note that the underlined operator $\hat l_0$ in (\ref{VirC}) is the grading operator, i.e. it commutes with any operator $\hat O^{(k)}$ of the grading $k$ as
\be
[\hat l_0,\hat O^{(k)}]=k\hat O^{(k)}
\ee
This implies that
\be
\hat l_0e^{\hat O^{(k)}}=e^{\hat O^{(k)}}(\hat l_0+k\hat O^{(k)})
\ee
and the equation
\be\label{Weq}
(\hat l_0-k\hat O^{(k)})\cdot Z=0
\ee
is solved by
\be\label{W}
Z=e^{\hat O^{(k)}}\cdot 1
\ee
Such a form of solution is usually called $W$-representation.
In particular, since the first three terms at the l.h.s. of (\ref{wn}) have the same grading $1$ w.r.t. this operator, it implies that Eq.(\ref{wn}) has a solution (cf. with \cite[Eq.(64)]{IMM}, \cite[Eq.(8)]{MM})
\be
Z_{N_1,N_2}\{p_k\} = \exp\Big(\hat w_{-1} + (N_1+N_2)\hat l_{-1} +N_1 N_2 p_1\Big)\cdot 1
\ee

Equation (\ref{wn}) can also be solved iteratively.
Let us again, similarly to (\ref{chaexpH}) denote the coefficients of expansion through $c_R$:
\be\label{chaexpC}
Z_{N_1,N_2}\{p\} = \sum_R c_R\cdot \s_R\{p\}
\ee
Then, one can use Eqs.(28) from \cite{MMMR1},
\be
\hat l_{-1}\chi_R=\sum_{R+\Box} (j_\Box-i_\Box) \chi_{R+\Box}\nn\\
\hat l_{0}\chi_R=|R|\cdot\chi_R\nn\\
\hat w_{-1}\chi_R=\sum_{R+\Box} (j_\Box-i_\Box)^2 \chi_{R+\Box}
\ee
in order to get from (\ref{wn}) that
\begin{equation}\label{rec}
|R|c_R=\sum_{R-\Box} (N_1+j_\Box-i_\Box)(N_2+j_\Box-i_\Box) c_{R-\Box}
\end{equation}
\\\\
As we explain in the next subsection, these recurrent equations uniquely define $c_R$.

\subsection{\boxed{\text{ Single equation is enough!}}}

Let us demonstrate that the solution to (\ref{rec}) is, indeed, unique. We start from symmetric representations,
\begin{equation}
    r c_{[r]}=(N_1-r-1)(N_2-r_1) c_{[r-1]}
\end{equation}
which define a recursion which is easily solved.
Similarly, the two-row partitions give rise to
\begin{equation}
    (r+1)c_{[r,1]}=(N_1+r-1)(N_2+r-1) c_{[r-1,1]}+(N_1-1)(N_2-1)c_{[r]}
\end{equation}
which again defines a simple recursion which can be solved:
\begin{equation}
    c_{[r,1]}=\dfrac{\chi_{[r,1]}(N_1)\chi_{[r,1]}(N_2)}{\chi_{[r,1]}}
\end{equation}
Further recursion
\begin{equation}
    (r+2)c_{[r,2]}=(N+r-1)(N+2-1) c_{[r-1,2]}+(N_1 N_2) c_{[r,1]}
\end{equation}
defines a similar recursion, with different coefficients, etc.
\\\\
Similarly, for any $R$, one can recursively use the equation until there is nothing left of the diagram $R$, and one is left with  $c_{[1]}= N_1 N_2 c_{[ \ ]} =N_1 N_2$. Hence, the recursion unambiguously determines $c_R$ for any $R$.

\subsection{How to construct the single equation\label{recipe}}

Let us discuss how one could guess an equation like Eq.(\ref{seqC}) that provides the necessary single equation. Notice that, in the Virasoro algebra (\ref{VirC}), all the terms but the last one have the same grading $-n$ and only the last one has the grading $-n-1$. What we do in Eq.(\ref{seqC}) is transforming this last term to the constraint $\hat l_0$ (see (\ref{wn})) of zero grading, which produces in the recursion relations (\ref{rec}) the l.h.s., i.e. the term $|R|c_R$, while all other terms become of grading 1 (so that one can use formulas (\ref{Weq})-(\ref{W})). In fact, this is the general rule of thumb: one has to make from the terms of non-standard grading (one could say, from the terms with dimensional coefficients) the $\hat l_0$ operator. This produces recursion relations that express $c_R$ through $c_{R'}$ with $|R'|<|R|$ only. We shall see in examples of the next section that this recipe always works and leads to $W$-representations like (\ref{W}).

\section{New approach: more examples}

\subsection{Gaussian Hermitian model}

As we already announced, the single equation for the Hermitian matrix model (\ref{Herm}) is
\be\label{seqH2}
\sum_{n\ge 1}p_nL^H_{n-2}Z_N\{p_k\}=0
\ee
(this coincides with the claim of \cite{Max2}).
Indeed, we can use our general recipe of sec.\ref{recipe}:
the sum should be split into a piece of definite grading $2$, and an additional
operator $\hat l_0$, which can be used to measure the grading:
\begin{equation}\label{eqH}
   \left(  \hat w_{-2} +2N\sum_{k=3} (k-2)p_k \dfrac{\partial }{\partial p_{k-2}} + N p_1^2+N^2 p_2- \underline{\hat l_0} \right)
   Z_N\{p_k\}=0
\end{equation}
Then, as follows from (\ref{W}),
\be
Z_N\{p_k\} = \exp\left({1\over 2}\hat w_{-2} +N\sum_{k=3} (k-2)p_k \dfrac{\partial }{\partial p_{k-2}}
+ {N p_1^2\over 2}+{N^2 p_2\over 2}\right)\cdot 1
\ee
is a solution to (\ref{eqH}), see also \cite{MSh,MM}.

(\ref{seqH2}) again trivially follows from the Virasoro constraints (\ref{vircon}), and,
in order to check that the solution is unique, we can use the formula (the simplest way to derive this formula is to use the fermionic representation \cite{MMMR3})
\begin{equation}
\hat w_{-2} \chi_R = \sum_{R+[2]} (j_\Box - i_\Box)(j_\Box-i_\Box+1) \chi_{R+[2]}
-\sum_{R+[1,1]} (j_\Box - i_\Box)(j_\Box-i_\Box-1) \chi_{R+[1,1]}
\end{equation}
and  get the recursion for the coefficients $c_R$ in (\ref{chaexpH})
\begin{equation}
   |R| c_R = \sum_{R-[2]}  (j_\Box - i_\Box+N)(j_\Box-i_\Box+1+N)c_{R-[2]} -\sum_{R-[1,1]}  (j_\Box - i_\Box+N)(j_\Box-i_\Box-1+N)c_{R-[1,1]}
\end{equation}
Now one can again do a recursion, by deleting $[2]$ and $[1,1]$ pieces, or, in combinatoric terms, 2-rim hooks or dominoes. One can immediately check that this recursion again has a unique solution \cite{MMMR3}.

\subsection{Cubic Kontsevich model}

Now we come to another type of matrix models, those depending on external matrix. Our first example is the simplest model of this kind, the Kontsevich model \cite{Kon}. Its partition function is given by the integral
\be\label{ZK}
Z_{K}(\Lambda)=\int dX \exp\left(-\Tr {X^3\over 3}-\Tr X^2\Lambda \right)
\ee
where $X$ is $N\times N$ Hermitian matrix, and the measure is normalized so that $\lim_{\Lambda\to\infty}Z_{K}(\Lambda)=1$. An important property of this integral \cite{Kon,GKM} is that it can be treated as a power series in $p_k:=\tr\Lambda^{-k}$, hence, we will use the notation $Z_K\{p_k\}$. Moreover, $Z_K\{p_k\}$ depends only on odd time-variables $p_{2k+1}$ \cite{GKM}, and the coefficients of this power series are just numbers, they do not depend on the size of matrix $X$ (if one considers large enough $N$, with only $p_{k<N}$ involved, see details in \cite{GKM}) .

This partition function satisfies the Virasoro constraints \cite{MMM,AMMP}
\be
\hat{ L}_n^K  Z_K\{p\} = 0, \ \ \ \ n\geq -1
\label{virK}
\ee
\be
\hat { L}_n^K :={1\over 2}\sum_{k} (k+2n)p_k\frac{\p}{\p p_{k+2n}} + {1\over 4}\sum_{a=1}^{2n-1} a(2n-a)\frac{\p^2}{\p p_a\p p_{2n-a}}
+ {p_1^2\over 4}\delta_{n,-1}+{1\over 16}\delta_{n,0}-\underline{(2n+3){\p\over\p p_{2n+3}}}
\label{VirK}
\ee
Here the sums over $k$ and $a$ run over odd numbers since $ Z_K\{p\}$ does not depend on $p_{2k}$. Now using our rule of thumb, we expect that the single equation in the Kontsevich model case is
\be\label{EqK}
\boxed{\sum_{n=1}p_{2n-1}\hat L^K_{n-2}Z_K\{p\}=0
}
\ee
which, in accordance with (\ref{W}), gives a solution \cite{Alex1}
\begin{equation}
Z_{K}(\Lambda)=\exp \left( \dfrac{1}{6} \sum_{k,l} (k+l-3)p_{k}p_{l}\dfrac{\partial}{\partial p_{k+l-3}}+\dfrac{1}{12}\sum_{k,l} p_{k}(k-l-3)l \dfrac{\partial^2}{\partial p_{k-l-3}\partial p_{l}}+\dfrac{1}{48}p_3+\dfrac{1}{12}p_1^3 \right)  \cdot 1
\end{equation}
since the grading is now equal to 3. The sums in the exponential run over odd integers.

Unambiguity of solution to (\ref{EqK}) can be checked by using the expansion of $Z_K\{p\}$ in, say, the Q Schur functions $Q_R\{p_k\}$ \cite{QKon}, which form a basis in the space of power series of odd time-variables $p_k$,
\be
Z_K\{p\}=\sum_R c_RQ_R\{p_k\}
\ee
One can use equally well any another basis.

For the coefficients $c_R$ of the expansion into the $Q$ Schur functions one again gets recursive equations, which are more involved \cite{MMMR3}. Simple examples are:
\begin{equation}
\begin{split}
&|R| c_{[r]}= -\dfrac{(2r-1)(2r-5)}{8}c_{[r-3]}\\
&|R| c_{[r-1,1]}=-\dfrac{1}{2} c_{[r-2]}+\dfrac{(2r-1)(2r-5)}{8}c_{[r-3,1]}
\end{split}
\end{equation}

\subsection{BGW model}

Our next example is the unitary matrix model depending on external matrix, which is usually called Br\'ezin-Gross-Witten (BGW) model \cite{BG,GW} and describes a generating function of Wilson averages in the lattice realization of the $2d$ gauge theory with the Wilson action. This generating function is given by the unitary matrix integral
\be
Z_{BGW}(M) = \int_{N\times N} dU \exp\left(\tr J^\dagger U +
\tr JU^\dagger\right)
\label{BGWorin}
\ee
where $dU$ is the Haar measure on unitary matrices, i.e. with the property $d(UV)=dU$ for a constant matrix $V$. Hence, the integral actually depends only on eigenvalues of Hermitian matrix $M:=JJ^\dagger$, i.e. on the variables of the form $\Tr (JJ^\dagger)^k$. We deal with this integral at large values of the eigenvalues of $M$, and normalize the measure so that $\lim_{M\to\infty}Z_{BGW}(M)=1$. Then, this integral can be treated as a power series in time-variables
\be\label{timesK}
p_k=\tr M^{-k/2}
\ee
and one can check by a direct (quite involved) calculation \cite{GKMU,AMMUC} that $Z_{BGW}\{p_k\}$
depends only on odd time-variables $p_{2k+1}$.

This property is quite similar to that of the Kontsevich model, and the partition function (\ref{BGWorin})
satisfies the Virasoro constraints \cite{GN,GKMU,AMMUC}
\be\label{virU}
\hat L^U_{n} Z_{BGW}\{p_k\} = 0,\ \ \ \ \ \ n\ge 0
\ee
\be\label{VirU}
\hat L^U_{n}={1\over 2}\sum_{k}(k+2n)p_k{\p\over\p p_{k+2n}}+{1\over 4}\sum_{{a,b}\atop{a+b=2n}}
ab{\p^2\over\p p_a\p p_b}+{\delta_{n,0}\over 16}-\underline{(2n+1){\p\over\p p_{2n+1}}}
\ee
where the sums run over odd $k$, $a$ and $b$. Hence, the differences with the Kontsevich model constraints are only in the grading shift, and in the lowest Virasoro constraint, much similar to the differences between the Virasoro constraints in the cases of complex and Hermitian matrix models. This means that the single equation should be the same as in the complex matrix model, (\ref{seqC}), but with the Virasoro constraints as in (\ref{VirU})
\be\label{seqU}
\boxed{
\sum_{n\ge 0} p_{2n+1}\hat{L}_{n}^UZ_{BGW}\{p_k\}=0
}
\ee
Since the grading is now equal to 1, one has a solution \cite{Alex2} (for the $W$-representation for the generalization of the BGW model, the antipolynomial GKM \cite{GKMU} see \cite{Alex3})
\be
Z_{BGW}\{p_k\}= \exp\left( \sum_{a,b} (a+b-1)p_{a}p_b{\p\over\p p_{a+b-1}}+
\frac{1}{4}\sum_{a,b}abp_{a+b+1}{\p^2\over\p p_a\p p_b} + \frac{p_1}{16}   \right)\cdot 1
\ee

\subsection{Generalized Kontsevich model with monomial potential}

A natural extension of the Kontsevich model is the generalized Kontsevich model (GKM) \cite{GKM}.
The monomial GKM is defined by the $N\times N$ Hermitian matrix integral \cite{GKM}
\be\label{GKM}
Z_{K_{n+1}}(\Lambda) :={\cal N}(\Lambda)\cdot \int  \exp\left(- {\Tr X^{n+1}\over n+1}+\Tr \Lambda^{n} X\right) dX
\ee
The potential in the exponent has an extremum at $X=\Lambda$,
and one expands around it in inverse powers of $\Lambda$, and choose the normalization factor
\be
{\cal N}(\Lambda):=
{\displaystyle{\exp\left({n\over n+1}\Tr \Lambda^{n+1}\right)}\over
\displaystyle{\exp\left(-\frac{1}{2}\sum_{a+b=n-1} \Tr \Lambda^a X \Lambda^b X\right) dX}}
\ee
This provides that $Z^{(n)}(\Lambda)$ can be understood as a formal power series in time-variables $p_k:=\Tr L^{-k}$, moreover, it does not depend on $p_{nk}$ \cite{GKM}.

The GKM partition function satisfies a more complicated set of the $W$-algebra constraints \cite{GKM}. We consider here only the case of quartic potential $n=3$, the extension to arbitrary $n$ being immediate. In this case, the partition function $Z_{K_4}\{p\}$ does not depend on the time-variables $p_{3k}$, and the set of constraints is
\be
\hat{ L}_n^{K_4}  Z_{K_4}\{p\} = 0, \ \ \ \ n\geq -1\nn\\
\nn\\
\hat{ W}_n^{K_4}  Z_{K_4}\{p\} = 0, \ \ \ \ n\geq -2
\label{virGKM}
\ee
\be
\hat { L}_n^{K_4} :={1\over 3}\sum_{k} (k+3n)p_k\frac{\p}{\p p_{k+3n}} + {1\over 6}\sum_{{a,b=1}\atop{a+b=3n}} ab\frac{\p^2}{\p p_a\p p_{b}}
+ {p_1p_2\over 3}\delta_{n,-1}+{1\over 9}\delta_{n,0}-\underline{(3n+4){\p\over\p p_{3n+4}}}
\nn\\
\hat { W}_n^{K_4} :=3\sum_{k,l=1}(k+l+3n)P_kP_l{\p\over\p p_{k+l+3n}}+3\sum_{k=1}\sum_{{a,b=1}\atop{a+b=k+3n}}baP_k{\p^2\over
\p p_a\p p_{b}}+\sum_{{a,b,c=1}\atop{a+b+c=3n}}abc{\p^3\over\p p_a\p p_b\p p_{c}}+\nn\\
+\sum_{{a,b,c=1}\atop{a+b+c=-3n}}P_aP_bP_{c}
\label{VirGKM}
\ee
where $P_k:=p_k-\underline{3\cdot\delta_{k,4}}$, and $a,b,c,k,l$ in the sums are not divisible by 3.

\bigskip

The underlined terms with non-standard grading are: the last term in the Virasoro generators, and terms containing the shift of $p_4$ in $P_4$. One can easily see that all terms that contain only linear derivative are summed up into $\hat l_0$ in the combinations $\sum_{n=1}p_{3n-1}\hat { W}_{n-3}^{(3)}$ and $\sum_{n=1}p_{3n-2}\hat { L}_{n-2}^{(3)}$. Hence, following our rule, we expect that the single equation generating the unique solution to  the constraints (\ref{virGKM}) is a sum of these two sums. In fact, the relative coefficient of these sums can be chosen almost arbitrary (only in the case of some positive rational coefficient, the soluton to this single equation is ambiguous) . The most convenient choice is -27 so that we finally choose
\be\label{GKMeq}
\boxed{\sum_{n=1}p_{3n-1}\hat { W}_{n-3}^{K_4}Z_{K_4}\{p\}-27
\sum_{n=1}p_{3n-2}\hat { L}_{n-2}^{K_4}Z_{K_4}\{p\}=0
}
\ee
for a single equation, which determines an unambiguous solution. This, indeed, works.

Let us note that the l.h.s. of (\ref{GKMeq}) is a sum of three operators of gradings 0, 4 and 8, hence, constructing a $W$-representation in this case is not that immediate. Note also that the zero grading term becomes the grading operator $\hat l_0=\sum_k p_k{\p\over\p p_k}$ exactly with the chosen coefficient -27, otherwise the zero grading term looks more involved, see (\ref{65}).

The relevant character expansion is in terms of the generalized $Q$-functions \cite{QGKM},
\begin{equation}
Z= \sum_R c_R Q^{(3)}_R
\end{equation}
Then equation \eqref{GKMeq} also translates into recursive equations for $c_R$. For example,
\begin{equation}\label{Q3rec}
\begin{split}
\dfrac{13}{9} c_{[4]}&=\frac{\left(2+2 i \sqrt{3}\right) \sqrt{18-6 i
   \sqrt{3}}}{\sqrt{3}+i} c_{[ \ ]}
   \\
   \dfrac{13}{9} c_{[3,1]}&=-\frac{9-3 i \sqrt{3}}{1+\sqrt[3]{-1}} c_{[ \ ]}
   \\
   \dfrac{16}{9} c_{[5]}&=30 c_{[1]}
   \\
   \dfrac{16}{9}
   c_{[4,1]}&=-\frac{6 \left(3-4 i
   \sqrt{3}\right)}{\left(1+\sqrt[3]{-1}\right)
   \sqrt{1-(-1)^{2/3}}} c_{[1]}
\end{split}
\end{equation}

Let us also mention a peculiarity. If one keeps in (\ref{GKMeq}) a generic coefficient $\alpha$ in front of the second term instead of setting $\alpha=-27$ as above, then the r.h.s of \eqref{Q3rec} becomes non-diagonal in $c_R$ and the generic equation is of the form
\begin{equation}\label{65}
\sum_{|P|=|R|} \xi_{R,P}(\alpha) c_P = \sum_{|P'|=|R|-4} \xi_{R,P',4}c_{P'} +\sum_{|P'|=|R|-8} \xi_{R,P',8}c_{P'}
\end{equation}
It appears that for certain values of $\alpha$ the determinant of $\xi_{R,P}$ vanishes, and the equations become degenerate allowing additional solution:
\be
|R|&=&4 ,\ \det\left(\xi_{R,P}\right) =-3456 (\alpha-27) \alpha ^2\nn\\
d |R|&=&5 , \  \det\left(\xi_{R,P}\right)=10125 (\alpha -108) (\alpha -18) \alpha ^2
\ee

\section{Conclusion}

In this letter, we reported a spectacular property of  matrix models:
their exact (non-perturbative) partition functions are unambiguously determined
by a {\it single} equation, which appears to be a $w$-substitute of the string equation.
It is still a question how universal this property is, but we explicitly demonstrated
that it holds for all basic models: rectangular complex, Hermitian, unitary, Kontsevich, and generalized Kontsevich models.
Moreover, as we explained, the relevant $w$-operators in all these cases but the last one has to reproduce the $W$-representations of the corresponding partition functions. This is, indeed, the case.

There are plenty of questions raised by this result.
Perhaps, the main one is that usually we have a {\it pair} of constraints:
either a pair of Virasoro generators, or a string equation acting on a restricted
space of KP-Toda $\tau$-functions.
How one can reduce a pair to just a single one?
Another essential question is how one can cook up a $W$-representation of the GKM partition function \cite{GKM} (like attempted in \cite{Zhou}) from the single equation formalism developed in this paper.
At last, there is a problem of constructing explicit solutions to the single equations using expansions of the corresponding partition functions in a basis of properly chosen symmetric functions. We return to all these questions elsewhere.

\section*{Acknowledgements}

Our work was supported in part by RFBR and NSFB according to the research project number 19-51-18006 (A.Mir., A.Mor.), by RFBR and TUBITAK, project number 21-51-46010 (A.Mir., A.Mor.), by RFBR and MOST, project number 21-52-52004 (A.Mir., A.Mor., V.Mish.). The work was also partly funded by the grant of the Foundation for the Advancement of Theoretical Physics ``BASIS" (A.Mir.), by RFBR grants 20-01-00644 (V.Mish) and 19-02-00815 (A.Mor.).
R~.R. was supported in part by FNI/BG-RU-2018/246, BNSF Grants H-28/5 and DN-18/1.

\section*{Appendix}

In this Appendix, we derive the determinant formula (\ref{detF}),
\begin{multline}  \label{our-det-1}
	\det_{1\le i,j\le N}{\delta^{(2)}_{N-1+R_i-i+j}\cdot(N-{2}+R_i-i+j)!!} =(-1)^{\frac{1}{2}N(N-1)} \prod_{i,j\in R}(N-i+j)\cdot\prod_{i=1}^{N-1}i!\cdot\det_{1\le i,j\le N}{\delta^{(2)}_{R_i-i+j}\over (R_i-i+j)!!}
\end{multline}
where $\delta^{(2)}_k:=(1+(-1)^k)/2$.

To compute/compare the two sides of \eqref{our-det-1} let us assume, without loss of generality, that $N$ is \textit{even} and consider the left hand side. The row elements of the left matrix are with alternating zeroes and manipulating rows and columns we can separate zeros from non-vanishing part making the matrix block diagonal.
\\\\\
We basically separate pieces for which $R_i-i$ is even into a matrix $A_K^L$, and those with $R_i-i$ odd, into $B_M^L$ Here $K,M$ are sizes of the corresponding block matrices, with $K+M=N$. For the right hand side, we do exactly the same separating into even and odd $R_i-i$. It is clear that the corresponding matrices are of the same size, hence our formula can be written in the form:
\be  \label{mat-1}
\operatorname{det}\begin{pmatrix} A_K^L & 0 \\ 0& B_M^L \end{pmatrix} = \xi \det \begin{pmatrix} A_K^R & 0 \\ 0 & B_M^R\end{pmatrix}
\ee
where the $N/2\times K$ matrix $A_K$ is given by
\begin{equation}
	A_K^L=
	\begin{pmatrix}
		{\color{red}(N-1+R_{i_1}-i_1)!!} & (N+1+R_{i_1}-i_1)!! & \cdots & (2N-1+R_{i_1}-i_1)!!  \\
		{\color{red}(N-1+R_{i_2}-i_2)!!} & (N+1+R_{i_2}-i_2)!! & \cdots & (2N-1+R_{i_2}-i_2)!! \\
		\vdots & \vdots & \vdots & \vdots  \\
		{\color{red}(N-1+R_{i_K}-i_K)!!} & (N+1+R_{i_K}-i_K)!! & \cdots & (2N-1+R_{i_K}-i_K)!!
	\end{pmatrix}
\end{equation}
and the $N/2\times M$ matrix $B_M^L$ reads off
\be
B_M^L=
\begin{pmatrix}
	{\color{blue}(N+R_{l_1}-l_1)!!} & (N+2+R_{l_1}-l_1)!! & \cdots & (2N-2+R_{l_1}-l_1)!! \\
	{\color{blue}(N+R_{l_2}-l_2)!!} & (N+2+R_{l_2}-l_2)!! & \cdots & (2N-2+R_{l_2}-l_2)!! \\
	\vdots & \vdots & \vdots & \vdots \\
	{\color{blue}(N+R_{l_M}-l_M)!!} & (N+2+R_{l_M}-l_M)!! & \cdots & (2N-2+R_{l_M}-l_M)!!
\end{pmatrix}
\ee
The right hand side matrices are written down below.
\\\\
We also need to carefully keep track of the signs appearing after manipulating rows and columns. First, we reorder rows of the both matrices: we just put even $R_i-i$ at the top, and odd ones at the bottom . The sign factor appearing from such manipulations is clearly the same on the left and the right. Reordering of columns in the right hand side differs by an additional pairwise permutation of adjacent elements, hence the relative sign is $(-1)^{N/2}$, which cancels with $(-1)^{\frac{N(N-1)}{2}}$ since we assumed $N$ is even. For odd $N$, the relative sign after column manipulations is $(-1)^{\left(N-1\right)/2}$ and cancels again.
\\\\
Therefore the prefactor in \eqref{mat-1} is
\begin{align}
	\xi=\prod_{i,j\in R}(N-i+j)\prod_{i=1}^{N-1}i! =  \prod_{s=1}^{N}(N+R_{s}-s)!\prod_{i=1}^{N-1}{i!\over (N-i)!}= \prod_{s=1}^{N} {\color{red}(N-1+R_{s}-s)!!}{\color{blue}(N+R_{s}-s)!!}
\end{align}

Note that the matrix $A_K^L$ contains elements of the form $(2n-1)!!$, the matrix $B_M^L$ has elements of the form $(2n)!!$ while  $A_K^R$ has elements of the form $(2n)!!$ but those in $B_M^R$ are of the form $(2n-1)!!$.

Pulling out the red elements in $A_K^L$ we find
\begin{align}
	A_K^L & = \prod_{k=1}^{K}{\color{red}(N-1+R_{i_k}-i_k)!!}
	\begin{pmatrix}
		1 & \frac{(N+1+R_{i_1}-i_1)!!}{{\color{red}(N-1+R_{i_1}-i_1)!!}} & \cdots & \frac{(2N-1+R_{i_1}-i_1)!!}{{\color{red}(N-1+R_{i_1}-i_1)!!}}  \\
		1 & \frac{(N+1+R_{i_2}-i_2)!!}{{\color{red}(N-1+R_{i_2}-i_2)!!}} & \cdots & \frac{(2N-1+R_{i_2}-i_2)!!}{{\color{red}(N-1+R_{i_2}-i_2)!!}} \\
		\vdots & \vdots & \vdots & \vdots  \\
		1 & \frac{(N+1+R_{i_K}-i_K)!!}{{\color{red}(N-1+R_{i_K}-i_K)!!}} & \cdots & \frac{(2N-1+R_{i_K}-i_K)!!}{{\color{red}(N-1+R_{i_K}-i_K)!!}}
	\end{pmatrix} \nonumber \\
	& = \prod\limits_{k=1}^{K}{\color{red}(N-1+R_{i_k}-i_k)!!}
	\begin{pmatrix}
		1 & (N+1+R_{i_1}-i_1) & \cdots & \prod\limits_{s=1}^{\frac{N}{2}}(N-1+R_{i_1}-i_1+2s)  \\
		1 & (N+1+R_{i_2}-i_2) & \cdots & \prod\limits_{s=1}^{\frac{N}{2}}(N-1+R_{i_2}-i_2+2s)  \\
		\vdots & \vdots & \vdots & \vdots  \\
		1 & (N+1+R_{i_K}-i_K) & \cdots & \prod\limits_{s=1}^{\frac{N}{2}}(R_{i_K}-i_K+2s)
	\end{pmatrix}
\end{align}
Under the determinant we can add and subtract columns without changing it.
Note that $p^{th}$ column contains polynomials in $(R_{i_p}-i_p)$ of order $p-1$. Multiplying first column by $N+1$ and subtracting it from the second one we find first order monomial $(R_{i_p}-i_p)$. Manipulating third column by multiplying first and second column appropriately and subtracting them from the third one we find for the third column monomial $(R_{i_p}-i_p)^2$. Proceeding in this way we can eliminate all the dependence on $N$ ending up with
\be
A_K^L = \prod\limits_{k=1}^{K}{\color{red}(N-1+R_{i_k}-i_k)!!}
\begin{pmatrix}
	1 & (R_{i_1}-i_1) & \cdots & (R_{i_1}-i_1)^{\frac{N}{2}}  \\
	1 & (R_{i_2}-i_2) & \cdots & (R_{i_2}-i_2)^{\frac{N}{2}}  \\
	\vdots & \vdots & \vdots & \vdots  \\
	1 & (R_{i_K}-i_K) & \cdots & (N-1+R_{i_K}-i_K)^{\frac{N}{2}}
\end{pmatrix}
\ee

Analogously, for $B_M^L$ we can write
\begin{align}
	B_M^L
	& = \prod_{s=1}^{M}{\color{blue}(N+R_{i_s}-i_s)!!}
	\begin{pmatrix}
		1 & \frac{(N+2+R_{l_1}-l_1)!!}{{\color{blue}(N+R_{l_1}-l_1)!!}} & \cdots & \frac{(2N-2+R_{l_1}-l_1)!!}{{\color{blue}(N+R_{l_1}-l_1)!!}}  \\
		1 & \frac{(N+2+R_{l_2}-i_2)!!}{{\color{blue}(N+R_{i_2}-i_2)!!}} & \cdots & \frac{(2N-2+R_{l_2}-l_2)!!}{{\color{blue}(N+R_{i_2}-i_2)!!}} \\
		\vdots & \vdots & \vdots & \vdots  \\
		1 & \frac{(N+2+R_{l_M}-l_M)!!}{{\color{blue}(N+R_{l_M}-l_M)!!}} & \cdots & \frac{(2N-2+R_{l_M}-l_M)!!}{{\color{blue}(N+R_{l_M}-l_M)!!}}
	\end{pmatrix} \nonumber \\
	& = \prod_{s=1}^{M}{\color{blue}(N+R_{i_s}-i_s)!!}
	\begin{pmatrix}
		1 & (N+2+R_{l_1}-l_1) & \cdots & \prod\limits_{s=1}^{\frac{N}{2}}(N+R_{i_1}-i_1+2s) \\
		1 & (N+2+R_{l_2}-l_2) & \cdots & \prod\limits_{s=1}^{\frac{N}{2}}(N+R_{i_2}-i_2+2s) \\
		\vdots & \vdots & \vdots & \vdots  \\
		1 & (N+2+R_{l_M}-l_M) & \cdots & \prod\limits_{s=1}^{\frac{N}{2}}(N+R_{i_M}-i_M+2s)
	\end{pmatrix} \nonumber
\end{align}
\begin{align}
	& = \prod_{s=1}^{M}{\color{blue}(N+R_{i_s}-i_s)!!}
	\begin{pmatrix}
		1 & (R_{l_1}-l_1) & \cdots & (R_{i_1}-i_1)^{\frac{N}{2}} \\
		1 & (R_{l_2}-l_2) & \cdots & (R_{i_2}-i_2)^{\frac{N}{2}} \\
		\vdots & \vdots & \vdots & \vdots  \\
		1 & (R_{l_M}-l_M) & \cdots & (R_{i_M}-i_M)^{\frac{N}{2}}
	\end{pmatrix}
\end{align}

Let us focus on the matrix on the \textbf{right hand side} corresponding to $A_K^L$
\begin{align}
	A_K^R & =
	\begin{pmatrix}
		\frac{1}{(R_{i_1}-i_1+2)!!} & \frac{1}{(R_{i_1}-i_1+4)!!} & \cdots & \frac{1}{(R_{i_1}-i_1+N)!! } \\
		\frac{1}{(R_{i_2}-i_2+2)!!} & \frac{1}{(R_{i_2}-i_2+4)!!} & \cdots & \frac{1}{(R_{i_2}-i_2+N)!!} \\
		\vdots & \vdots & \vdots & \vdots  \\
		\frac{1}{(R_{i_K}-i_K+2)!!} & \frac{1}{(N+1+R_{i_K}-i_K+4)!!} & \cdots & \frac{1}{(2N-1+R_{i_K}-i_K+N)!!}
	\end{pmatrix} \nonumber \\
	& = {\color{blue}\prod\limits_{s=1}^M \frac{1}{(R_{i_s}-i_s+N)!!}  }
	\begin{pmatrix}
		\prod\limits_{s=2}^{\frac{N}{2}}(R_{i_K}-i_K+2s) & \dots & (R_{i_1}-i_1+4) & 1 \\
		\prod\limits_{s=2}^{\frac{N}{2}}(R_{i_K}-i_K+2s) & \dots & (R_{i_1}-i_1+4) & 1 \\
		\vdots & \vdots & \vdots & \vdots  \\
		\prod\limits_{s=2}^{\frac{N}{2}}(R_{i_K}-i_K+2s) & \dots & (R_{i_1}-i_1+4) & 1 \\
	\end{pmatrix}
\end{align}
We can manipulate the matrix in the very same way as we did with $A_K^L$. The result is an equivalent equivalent for the determinant of matrix containing only powers of $(R_i-i)$, namely
\be
A_K^R = {\color{blue}\prod\limits_{s=1}^M \frac{1}{(R_{i_s}-i_s+N)!!}  } \begin{pmatrix}
	1 & (R_{l_1}-l_1) & \cdots & (R_{i_1}-i_1)^{\frac{N}{2}} \\
	1 & (R_{l_2}-l_2) & \cdots & (R_{i_2}-i_2)^{\frac{N}{2}} \\
	\vdots & \vdots & \vdots & \vdots  \\
	1 & (R_{l_M}-l_M) & \cdots & (R_{i_M}-i_M)^{\frac{N}{2}}
\end{pmatrix},
\ee
where we moved the last column to the first position etc.

The two sides of the equality we want to prove, restricted to the matrices $A_K^L$ and $A_K^R$ become
\begin{multline}
	\prod\limits_{k=1}^{K}{\color{red}(N-1+R_{i_k}-i_k)!!}
	\begin{pmatrix}
		1 & (R_{i_1}-i_1) & \cdots & (R_{i_1}-i_1)^{\frac{N}{2}}  \\
		1 & (R_{i_2}-i_2) & \cdots & (R_{i_2}-i_2)^{\frac{N}{2}}  \\
		\vdots & \vdots & \vdots & \vdots  \\
		1 & (R_{i_K}-i_K) & \cdots & (N-1+R_{i_K}-i_K)^{\frac{N}{2}}
	\end{pmatrix} \\
	= \prod_{s=1}^{N} {\color{red}(N-1+R_{s}-s)!!}{\color{blue}(N+R_{s}-s)!!}.
	{\color{blue}\prod\limits_{s=1}^M \frac{1}{(R_{i_s}-i_s+N)!!}  }
	\begin{pmatrix}
		1 & (R_{l_1}-l_1) & \cdots & (R_{i_1}-i_1)^{\frac{N}{2}} \\
		1 & (R_{l_2}-l_2) & \cdots & (R_{i_2}-i_2)^{\frac{N}{2}} \\
		\vdots & \vdots & \vdots & \vdots  \\
		1 & (R_{l_M}-l_M) & \cdots & (R_{i_M}-i_M)^{\frac{N}{2}}
	\end{pmatrix}
\end{multline}
The very same procedure can be applied to the other matrices, $B_M^L$ and $B_M^R$.
Cancelation of the corresponding colored factors on the both sides proves the formula.

\end{document}